\documentclass[aps,prd,twocolumn,groupedaddress,
a4paper,floats]{revtex4}

\usepackage{graphicx}
\usepackage{amssymb,latexsym}
\usepackage{longtable}

\begin{document}

\title{Triality between Inflation, Cyclic and Phantom Cosmologies}

\author{James E. Lidsey}
\affiliation{Astronomy Unit, 
School of Mathematical Sciences, Queen Mary, University of
London, Mile End Road, E1 4NS, UK}


\begin{abstract}
It is shown that any spatially flat and isotropic universe undergoing 
accelerated expansion driven by a self-interacting scalar field can be 
directly related to a contracting, decelerating cosmology. 
The duality is made manifest by expressing the scale factor 
and Hubble parameter as functions of the scalar field and 
simultaneously interchanging these two quantities.  
The decelerating universe can be twinned with a cosmology sourced 
by a phantom scalar field by inverting the scale factor 
and leaving the Hubble parameter invariant. The accelerating model 
can be related to the same phantom universe by identifying the scale 
factor with the inverse of the Hubble parameter.   
The duality between accelerating and decelerating backgrounds
can be extended to spatially curved cosmologies and models 
containing perfect fluids. A similar triality and associated scale 
factor duality is found in the Randall-Sundrum type II braneworld scenario.
\end{abstract}


\maketitle

Spatially isotropic and homogeneous 
Friedmann--Robertson--Walker (FRW) universes containing a 
scalar field, $\phi$, that is minimally coupled to Einstein gravity and 
self--interacting through a potential, $V(\phi )$, provide an
important framework for modern cosmology. If the potential is positive 
and sufficiently flat, such a field can drive an epoch of rapid, 
accelerated, inflationary expansion. (For a review, see, e.g., 
Ref. \cite{lidlyth}). On the other hand, the universe 
undergoes a phase of slow, 
decelerated contraction if the potential is steep and negative. This 
latter type of potential arises in the ekpyrotic/cyclic cosmological scenario 
\cite{cyclic,null} based  
on brane collisions in heterotic M--theory \cite{witten}, 
where the scalar field is the 
moduli field parametrizing the separation between the branes. 

In a collapsing, spatially flat FRW cosmology, the null energy 
condition $\rho +p \ge 0$ 
must be violated if the bounce between the collapsing and 
expanding phases is to be non--singular \cite{null}. 
Violation of this condition is possible in cosmologies 
sourced by `phantom' matter with an equation of state $w = p/\rho < -1$.  
A scalar field with negative kinetic energy is one example of 
this type of matter
\cite{caldwell,cht}.
Although phantom matter leads to instabilities in any low energy theory 
\cite{cline}, there has recently been considerable 
interest in phantom cosmologies. This development has been motivated by
observations of type Ia supernovae \cite{sn1}: 
dark energy that violates the null energy condition 
is consistent with the data \cite{ck}. 
String theory provides further 
motivation for considering phantom matter \cite{frampton}.  
Moreover, it was 
recently shown that a scale--invariant perturbation spectrum can be 
generated in a collapsing model sourced by both a standard  
and a phantom scalar field \cite{aw}.
Finally, a universe dominated by a phantom scalar field undergoes 
superinflationary expansion, where the Hubble parameter increases with time
and early universe models driven by such a field have been discussed 
\cite{piao}. 

The comoving Hubble scale decreases during a phase of 
accelerated expansion or decelerated contraction. 
This implies that inflation and cyclic cosmologies can, 
in principle, generate density perturbations on scales larger than the Hubble 
radius at the epoch of decoupling \cite{gkst}. 
Evidence for the existence of large--scale 
perturbations is found in the anti--correlation of the temperature 
anisotropy and polarization maps detected by the WMAP satellite 
\cite{white,kogut}. Although the mechanisms 
that generate the perturbations in the two scenarios are radically 
different, a surprising duality between the spectra 
has recently been uncovered \cite{kst,bst}:  
for a given density perturbation spectrum generated from inflationary 
expansion with a spectral index $n_S$, there exists a contracting model that 
produces a spectrum with the same value of $n_S$. 
More specifically, the spectral index is given by 
\begin{equation}
\label{tilt}
n_S -1 \approx - \frac{2}{(1-\epsilon)^2} \left[ 
\epsilon - \frac{(1-\epsilon^2 )}{2} \left( 
\frac{d \ln \epsilon}{d {\cal{N}}} \right) \right]  ,
\end{equation}
where $\epsilon \equiv -\dot{H}/H^2$, ${\cal{N}} 
= \ln [a_e H_e/ aH ]$, $H$ is the Hubble parameter, $a$ is the scale 
factor, a subscript `e' denotes values at the 
end of the inflationary or contracting phase and  
higher--order derivatives of the form 
$(d \ln \epsilon /d {\cal{N}} )^2$ and 
$d^2 \ln  \epsilon / d {\cal{N}}^2$ have been neglected. 
Eq. (\ref{tilt}) is invariant 
under the duality $\epsilon \rightarrow 1/\epsilon$ and, consequently, 
identical, nearly scale--invariant
perturbation spectra are generated by phases of rapid, 
accelerated expansion $(\epsilon \ll 1)$ and slow, decelerated contraction
$(\epsilon \gg 1)$ \cite{kst,bst}. 
In the limit where $\epsilon = {\rm constant}$, 
it can be further shown that this duality is exact \cite{bst}. 

The duality between inflation and cyclic cosmology 
suggests there may be some formal correspondence
between the two scenarios at the level of the effective, 
four--dimensional field equations.  
One method of generating inflating cosmologies 
from non--inflating backgrounds is to employ the form--invariance
properties of the FRW field equations 
\cite{chimento,acjl}. Motivated by the above developments, 
the purpose of the present work is to show that 
inflationary, cyclic and phantom 
FRW cosmologies can be related through a series 
of symmetry transformations. This triality of transformations 
maps a given cosmological solution parametrized by a scale factor and 
scalar field potential onto other solutions sourced by fields with 
different interaction potentials. 

We first consider the spatially flat FRW background. 
The Friedmann and scalar field equations 
are given by 
\begin{eqnarray}
\label{friedmann}
3H^2 = \frac{\ell}{2} \dot{\phi}^2 +V \\
\label{scalarfield}
\ddot{\phi} + 3 H \dot{\phi} + \ell V' =0  ,
\end{eqnarray}
where $\ell =\pm 1$ for conventional and phantom matter, 
respectively, a prime denotes differentiation with 
respect to the scalar field and units are chosen such that $c = 8\pi G =1$.
The energy density is given by $\rho = \ell \dot{\phi}^2 /2 +V$. 
Eq. (\ref{scalarfield}) can be expressed in the form
\begin{equation}
\label{dotrho}
\dot{\rho} = -3 \ell H \dot{\phi}^2  
\end{equation}
or, equivalently, as
\begin{equation}
\label{dotH}
\dot{H} = -\frac{ \ell}{2} 
\dot{\phi}^2
\end{equation}
and Eq. (\ref{dotH}) can be rewritten in the form of 
a time--dependent harmonic oscillator: 
\begin{equation}
\label{ho}
\frac{d^2 b}{d \tau^2} - \frac{\ell}{2} \left( \frac{d\phi}{d \tau} 
\right)^2 b =0  ,
\end{equation}
where $b \equiv a^{-1}$ and $\tau \equiv \int^t dt/a(t)$ defines 
conformal time. 

If a particular solution, $b_1 (\tau)$, to Eq. 
(\ref{ho}) is known, a second, linearly independent solution is given 
in terms of a quadrature of the first: 
$b_2  = b_1 \int^{\tau} d \tau /b_1^2$. 
This indicates that for any given expanding, accelerating 
universe, there exists a dual cosmology 
where the functional form of the scalar field 
(when expressed in terms of conformal time) 
is invariant. In particular, for the {\em ansatz} 
$\phi = [\sqrt{2q} /(1-q)] \ln \tau$, where $q$ is a constant and $\ell =1$, 
Eq. (\ref{ho}) simplifies to 
\begin{equation}
\label{hop}
\frac{d^2b}{d\tau^2} - \frac{q}{(1-q)^2}
\frac{1}{\tau^2} b =0
\end{equation}
and admits the two linearly independent
solutions, $b_1 \propto \tau^{q/(q-1)}$ and 
$b_2 \propto \tau^{1/(1-q)}$. In terms of proper time, 
these correspond to 
$a \propto t^q$ and $a \propto t^{1/q}$, respectively, and  
the two branches are related by the 
transformation $q \rightarrow 1/q$. 
Indeed, Eq. (\ref{hop}) is invariant 
under this transformation. For $q> 1$,  
the time reversal of the second branch represents the decelerating, 
contracting solution of the cyclic 
scenario when the equation of 
state $\epsilon = 3(1+w)/2 = 1/q$ is constant \cite{bst}. 

Further insight may be gained by expressing 
Eqs. (\ref{friedmann}) and (\ref{scalarfield}) 
in the `Hamilton--Jacobi' form \cite{HJ,HJ1,HJ2}: 
\begin{eqnarray}
\label{HJ1}
V= 3H^2 -2\ell H'^2 \\
\label{HJ2}
H' = -\frac{\ell}{2} \dot{\phi}   ,
\end{eqnarray}
where the Hubble parameter is viewed as a function of the scalar field
and it is assumed that the field is a monotonically varying function of 
proper time. 
It follows from the definition of the Hubble parameter that 
\begin{equation}
\label{Hubbleprime}
a'H' = - \frac{\ell}{2} a H
\end{equation}
and integrating Eq. (\ref{Hubbleprime}) implies that
\begin{equation}
\label{aquad}
a (\phi) = \exp \left[ - \frac{\ell}{2} \int^{\phi} 
d \phi \frac{H}{H'} \right]  .
\end{equation}
For a particular functional form of the Hubble parameter, $H(\phi)$, 
the potential is given directly by Eq. (\ref{HJ1})
and the scale factor is determined 
in terms of the single quadrature (\ref{aquad}).
The time--dependence of the scalar field can be deduced by integrating 
Eq. (\ref{HJ2}): 
\begin{equation}
\label{integrateHJ2}
t= - \frac{\ell}{2}\int^{\phi} \frac{d\phi}{H'}  .
\end{equation}

We begin by discussing the duality that maps a conventional 
expanding, 
accelerating universe onto a decelerating background
($\ell = 1$). 
It is immediately apparent that Eq. (\ref{Hubbleprime}) is invariant 
under the simultaneous interchange $H(\phi)  \leftrightarrow a(\phi)$.
We therefore consider the solution $\{ H(\phi) , V(\phi) , a(\phi) \}$ for 
a standard scalar field cosmology
and a new {\em ansatz} for the Hubble parameter 
of the form $\tilde{H}(\phi) = a(\phi)$. 
It follows from Eq. (\ref{aquad}) that the new scale factor is given by 
\begin{equation}
\label{newscalequad}
\tilde{a}(\phi ) = \exp \left[ - \frac{1}{2} \int^{\phi} d\phi 
\frac{a}{a'} \right]  .
\end{equation}
However, since $a(\phi)$ is itself a solution to the field 
equations, it satisfies 
Eq. (\ref{Hubbleprime}). Hence, modulo an irrelevant 
constant of proportionality, the new scale factor 
is given by 
\begin{equation}
\label{duality1}
\tilde{a}(\phi) = H(\phi) , \qquad \tilde{H} (\phi) = a(\phi)
\end{equation}
and the new potential is given in terms of the old Hubble parameter by 
\begin{equation}
\label{newpotential}
\tilde{V} = \left( 3 - \frac{1}{2} \frac{H^2}{H'^2} \right) 
\exp \left[ - \int^{\phi} d\phi \frac{H}{H'} \right]   .
\end{equation}

The duality transformation (\ref{duality1}) inverts the Hubble--flow 
parameter, $\epsilon \equiv -\dot{H}/H^2$: 
\begin{equation}
\label{epsilon}
\epsilon = 2\frac{H'^2}{H^2} , \qquad \tilde{\epsilon} = \frac{1}{\epsilon}
\end{equation}
and an accelerating, expanding universe is therefore mapped onto 
a decelerating solution. 
The self--dual solution is the coasting cosmology, $a=t$. 
The dual cosmology is itself 
an expanding model since $\tilde{H} (\phi) = a (\phi ) \ge 0$. 
However, a time reversal leads immediately to a contracting solution.
Moreover, an accelerating, expanding 
cosmology with $\epsilon < 1/3$ is dual to a decelerating, contracting 
model with a negative potential. Indeed, 
it can be shown that the expanding solutions are stable to small 
perturbations if $\epsilon <1$ 
(corresponding to inflation driven by a positive potential) and 
contracting cosmologies are stable if $\tilde{\epsilon} > 3$ (corresponding to 
the cyclic scenario driven by a negative potential) \cite{bst,ewst}. 
We therefore refer to the dual 
solution $\tilde{a} (\phi)$ as the cyclic cosmology. 

To summarize thus far, inflating and cyclic cosmologies can 
be related by a simultaneous interchange of the Hubble parameter 
and the scale factor {\em when both are expressed as functions of the 
scalar field}. 
This generalizes the result of Ref. \cite{bst} for 
constant $\epsilon$ to that   
of arbitrary scalar field potentials, where  
$\epsilon$ is time--dependent. The dual potentials are related by a single 
quadrature involving the Hubble parameter. The comoving Hubble radius 
$[a (\phi) H (\phi)]^{-1}$ remains invariant under the duality and this implies  
that the derivative operator $d/d{\cal{N}} = - d/d\ln (aH )$ arising
in expression (\ref{tilt}) for the spectral index is also invariant. 

In the slow--roll inflationary limit, 
$3H^2 (\phi ) \approx V(\phi)$ and $\epsilon \approx 
V'^2/(2V^2) \ll 1$. It follows from Eq. (\ref{tilt}) 
that the spectral index of 
the perturbation spectrum generated during inflation is then given in terms 
of the potential by 
\begin{equation}
\label{spectralinflation}
n_S -1 \approx -3 \frac{V'^2}{V^2} + 2 \frac{V''}{V}  .
\end{equation}
Eq. (\ref{spectralinflation}) 
can be expressed in terms of the dual cyclic potential. In the 
slow--roll limit of inflation, Eq. (\ref{newpotential})
simplifies to 
\begin{equation}
\label{epot}
\tilde{V} \approx - 2 \frac{V^2}{V'^2} \exp \left[ - 2 \int^{\phi}
d \phi \frac{V}{V'} \right]  
\end{equation}
and differentiating Eq. (\ref{epot}) with respect to the 
scalar field implies that 
\begin{equation}
\label{epotdual}
\frac{\tilde{V}'}{\tilde{V}} \approx 2 \left( \frac{V'}{V} - \frac{V''}{V'} 
- \frac{V}{V'} \right)  .
\end{equation}
Since the third term on the right hand side 
of Eq. (\ref{epotdual}) dominates, it follows 
that $V'/V \approx -2 \tilde{V}/
\tilde{V}'$. (This approximation is equivalent to the 
assumption that both potentials are approximately exponential in form). 
Differentiation of this relation with 
respect to the scalar field and 
substitution of the result back into Eq. (\ref{spectralinflation})
then results in the spectral index of density perturbations
for the cyclic scenario \cite{gkst}: 
\begin{equation}
\label{spectralcyclic}
{n}_S -1 \approx  -4 \left( 1 + 
\frac{\tilde{V}^2}{\tilde{V}'^2} - 
\frac{\tilde{V}\tilde{V}''}{\tilde{V}'^2}
\right)   .
\end{equation}

We now consider the map between a standard scalar field 
cosmology $\{ H(\phi ), V(\phi ), a(\phi) \}$ and a phantom cosmology with 
$\ell =-1$. If the dual Hubble parameter is given by 
$\hat{H} = a(\phi)$, the phantom cosmology is determined by the
quadrature 
\begin{equation}
\label{phantomnew}
\hat{a} (\phi) = \exp \left[ \frac{1}{2} \int^{\phi} d \phi \frac{a}{a'}
\right]  .
\end{equation}
Since $a(\phi)$ represents a solution to the standard 
scalar field equations, it satisfies Eq. (\ref{Hubbleprime})
with $l=1$. 
Substituting Eq. (\ref{Hubbleprime}) into Eq. (\ref{phantomnew}) 
and integrating then implies that 
\begin{equation}
\label{phantomscale}
\hat{a} (\phi ) = 1/ H(\phi) , \qquad \hat{H}(\phi)  = a(\phi )
\end{equation}
and the phantom potential 
is given by 
\begin{equation}
\label{phantompotential}
\hat{V} = \left( 3 + \frac{1}{2} \frac{H^2}{H'^2} \right) 
\exp \left[ - \int^{\phi} d\phi \frac{H}{H'} \right]  .
\end{equation}
Thus, a positive or negative
potential is mapped onto a positive potential, whereas the sign of the field's 
kinetic energy is reversed. The transformation inverts and simultaneously
changes the sign of the Hubble--flow parameter,
$\hat{\epsilon} = -1/\epsilon$. 

The triality is completed by 
mapping the two dual cosmologies (\ref{duality1}) and (\ref{phantomscale}) 
directly onto one another. Comparison of Eqs. (\ref{duality1}) 
and (\ref{phantomscale}) immediately implies that the 
duality inverts the scale factors when each is expressed as 
a function of the scalar field: 
\begin{equation}
\label{invert}
\tilde{a} (\phi ) = \frac{1}{\hat{a} (\phi)}  , \qquad 
\tilde{H} (\phi ) = \hat{H} (\phi)   .
\end{equation}
This is similar to the scale factor duality 
of string cosmology \cite{sfd}, although in that case the field equations 
remain invariant whereas the transformation (\ref{invert}) relates solutions 
to field equations derived from different actions. (For reviews, see 
Refs. \cite{lwc,gv}.)
Scale factor dualities between standard and phantom cosmologies 
have been found previously 
in the case where the phantom matter is a perfect fluid \cite{dss} 
and a self--interacting scalar field \cite{cl}. In the latter case, however, 
the transformation involved a Wick rotation of the scalar field 
together with a change in sign of the Hubble parameter.   
The Hubble parameter is invariant in Eq. (\ref{invert}). 
Moreover, Eqs. (\ref{newpotential}) 
and (\ref{phantompotential}) imply that the two dual potentials
generated from the same inflationary 
background where $\epsilon \ll 1$  differ 
only by a sign, $\tilde{V} (\phi ) \approx - \hat{V} (\phi )$. 

As an illustrative example of the triality, 
consider the power law solution \cite{lm}: 
\begin{equation}
\label{powerlaw}
a= t^{2 /\lambda^2} , \qquad \phi = \frac{2}{\lambda} \ln t , 
\qquad V= \frac{2(6- \lambda^2)}{\lambda^4} e^{-\lambda \phi} ,
\end{equation}
where $\lambda^2<6$ is a constant. 
In terms of the scalar field, the Hubble parameter and 
scale factor are given by $H(\phi ) = (2/\lambda^2)e^{-\lambda\phi /2}$
and $a(\phi ) = e^{\phi / \lambda}$, respectively. 
Employing the duality transformation (\ref{duality1}) 
for $l=1$ and integrating 
Eq. (\ref{integrateHJ2}) implies that 
\begin{eqnarray}
\label{ekpower}
\tilde{a} = t^{\lambda^2 /2} , \quad \tilde{\phi} = -\lambda \ln t
\nonumber \\
\tilde{V} = \frac{\lambda^2}{4} \left( 3 \lambda^2 -2 
\right) e^{2 \phi / \lambda}  ,
\end{eqnarray}
where we have rescaled the time variable $t \rightarrow 
2t/\lambda^2$ without loss of generality. The time reversal of 
solution (\ref{ekpower}) 
is the canonical cyclic cosmology and represents a slowly collapsing 
universe for $\lambda \ll 1$ \cite{bst}. 

The duality (\ref{phantomscale}) maps the power law solution 
(\ref{powerlaw}) onto the corresponding phantom cosmology:
\begin{eqnarray}
\label{phantompower}
\hat{a} = (-t)^{-\lambda^2 /2}, \quad \hat{\phi} = - \lambda 
\ln  (-t) 
\nonumber \\
 \hat{V} = \frac{\lambda^2}{4} 
\left( 3 \lambda^2 +2 \right) e^{2 \phi / \lambda}  ,
\end{eqnarray} 
where the time variable is rescaled such that
$t \rightarrow 2t/\lambda^2$. Eq. (\ref{phantompower}) 
is the power law superinflationary model considered 
recently in Ref. \cite{piao}. 
The scale factors in Eqs. (\ref{ekpower}) and 
(\ref{phantompower}) are indeed the inverse of each other. 

Thus far, we have restricted the discussion to the spatially 
flat FRW models sourced by a single, self--interacting scalar 
field. It is of interest to investigate whether 
the dualities discussed above can be extended to more general backgrounds and 
matter sources. Of particular interest is the class of 
models containing both a scalar field and a perfect fluid with 
barotropic equation of state $p_{\rm mat}=(\gamma -1)\rho_{\rm mat}$, 
where $\gamma \le 2$ is a constant and $p_{\rm mat}$ and 
$\rho_{\rm mat}$ represent the pressure and energy density of the 
fluid, respectively. The matter 
can be interpreted as a phantom fluid for 
$\gamma < 0$. If the fields are uncoupled, energy--momentum
conservation implies that $\rho_{\rm mat}= 
Da^{-3\gamma}$, where $D$ is a constant. The 
Friedmann equation is then given by  
\begin{equation}
\label{curvedfriedmann} 
3H^2 = \rho + \frac{D}{a^{3\gamma}}
\end{equation}
and the case of $\gamma = 2/3$ may be interpreted 
as a positively (negatively) 
curved FRW universe if $D=-1$ $(D=+1)$. 

The scalar field equation, 
Eq. (\ref{dotrho}), can be expressed in
the form $\rho' = -3 \ell H \dot{\phi}$
if the field is a monotonically varying function of proper time.  
The definition of the Hubble parameter 
then implies that $9 \gamma H^2 = -\ell \chi' \rho' /\chi$, where $\chi 
\equiv a^{3 \gamma}$. Substituting this expression into Eq. 
(\ref{curvedfriedmann}) allows the
Friedmann equation to be expressed in the form \cite{HJ1}
\begin{equation}
\label{curveprime}
\rho' (\phi ) \chi' (\phi ) +3 \ell \gamma \rho (\phi ) \chi (\phi )  = 
- 3 \ell \gamma D  .
\end{equation}
 
The general solution to Eq. (\ref{curveprime}) 
can be expressed in terms of quadratures: 
\begin{eqnarray}
\chi (\phi ) = \exp \left[ -3\ell \gamma \int^{\phi} d\phi \frac{\rho}{\rho'}
\right] 
\nonumber \\
\label{curvequad}
\times \left[ \Pi - 3 \ell \gamma D \int^{\phi} d\phi \frac{1}{\rho'} \exp 
\left( 3 \ell \gamma \int^{\varphi} d \varphi \frac{\rho}{\rho'} \right)
\right]  ,
\end{eqnarray}
where $\Pi$ is an arbitrary integration constant. 
However, since 
Eq. (\ref{curveprime}) is invariant 
under the simultaneous interchange $\rho (\phi ) \leftrightarrow 
\chi (\phi )$, the general solution to Eq. (\ref{curveprime}) 
can also be expressed in the form
\begin{eqnarray}
\rho (\phi ) = \exp \left[ -3\ell \gamma \int^{\phi} d\phi \frac{\chi}{\chi'}
\right] 
\nonumber \\
\label{rhocurvequad}
\times \left[ \Pi - 3 \ell \gamma D \int^{\phi} d\phi \frac{1}{\chi'} \exp 
\left( 3 \ell \gamma \int^{\varphi} d \varphi \frac{\chi}{\chi'} \right)
\right]  .
\end{eqnarray}

If we now consider a solution 
$\{ \rho (\phi ) , \chi (\phi) \}$ for a standard scalar field 
cosmology $(\ell =1)$ and invoke a new {\em ansatz}  
$\tilde{\rho} (\phi ) = \chi (\phi)$, comparison of  
Eqs. (\ref{curvequad}) and (\ref{rhocurvequad}) implies 
that the dual cosmology is given by $\tilde{\chi} (\phi ) = \rho (\phi)$
if the equation of state of the fluid, $\gamma$, remains invariant.
It follows that  
\begin{equation}
\label{rhochi}
\tilde{\rho} (\phi ) = a^{3 \gamma} (\phi ) , \qquad 
\tilde{a} (\phi) = [\rho (\phi ) ]^{1 /3\gamma}
\end{equation}
and it may be verified directly that for $\gamma = 2/3$ 
the coasting solution, $a = t$, is self--dual 
under the transformation (\ref{rhochi}). We conclude, 
therefore, that the spatially flat duality  
between expanding, accelerating 
cosmologies and contracting, decelerating models may be extended to 
include spatial curvature and perfect fluid sources. 

In the spatially flat model, the phantom duality (\ref{phantomscale})
arises because Eq. (\ref{Hubbleprime}) is invariant under 
the simultaneous interchange $ H \rightarrow a$, 
$a \rightarrow 1/H$ and $\ell \rightarrow -\ell$. 
However, Eq. (\ref{curveprime}) is not invariant 
under $\rho \rightarrow \chi$, $\chi \rightarrow 1/\rho$ and 
$\ell \rightarrow -\ell$ when $D \ne 0$. 
On the other hand, a change in the sign of $\ell$ does leave 
Eq. (\ref{curveprime}) invariant if 
the sign of the barotropic index also changes, 
$\gamma \rightarrow -\gamma$. As a result, 
a standard scalar field/perfect fluid cosmology can be mapped 
onto a phantom model where both the scalar field and fluid 
are phantoms. The transformation is $\rho (\phi ) \leftrightarrow 
\chi (\phi)$, $\gamma \rightarrow -\gamma$ and $\ell \rightarrow -\ell$. 
The necessary change in $\gamma$ indicates that the phantom duality 
(\ref{phantomscale}) can not be extended to 
spatially curved FRW backgrounds. 

A further question of importance is whether similar dualities can be 
found in other cosmological scenarios developed from gravitational 
physics different to that of four--dimensional Einstein gravity.
A much studied model is the Randall--Sundrum type II (RSII) braneworld, 
where a co--dimension one brane 
is embedded in five--dimensional Anti--de 
Sitter (AdS) space \cite{RSII}. In this model, the Friedmann equation acquires 
a quadratic 
dependence on the energy density of matter confined to the brane
\cite{quad}: 
\begin{equation}
\label{RSfriedmann}
3H^2 = \rho +\frac{\rho^2}{2 \lambda}  ,
\end{equation}
where the brane tension, $\lambda$, has been tuned so that 
the effective four--dimensional cosmological constant vanishes. 
The equation of motion for 
a single scalar field confined to the brane
is given by Eq. (\ref{dotrho}). 

Eqs. (\ref{RSfriedmann}) and (\ref{dotrho}) can be written in an 
alternative form by defining the new variable \cite{hawlid}:
\begin{equation}
\label{defy}
y = \left( \frac{\rho}{\rho + 2 \lambda} \right)^{1/2}   .
\end{equation}
Substituting Eq. (\ref{defy}) into Eqs. (\ref{RSfriedmann}) 
and (\ref{dotrho}) then implies that 
\begin{eqnarray}
\label{defH}
H= \left( \frac{2 \lambda}{3} \right)^{1/2} \frac{y}{1-y^2}
\\
\label{dotphiRS}
\dot{\phi} = - \ell \left( \frac{8 \lambda}{3} \right)^{1/2} 
\frac{y'}{1-y^2}
\end{eqnarray}
and it follows from the definition of the Hubble parameter 
that \cite{hawlid}
\begin{equation}
\label{ay}
y'a' = -\frac{\ell}{2} ya    .
\end{equation}
Eq. (\ref{ay}) is invariant under the transformation 
$y(\phi ) \leftrightarrow a (\phi )$ and has 
an identical form to that of Eq. (\ref{Hubbleprime}). Consequently, 
the above triality for spatially flat FRW models based on Einstein 
gravity can be directly extended to the RSII braneworld scenario
by simply replacing the Hubble parameter with the variable $y(\phi )$.
We conclude, therefore, that there exists a triality in 
the RSII braneworld of the form
\begin{equation}
\label{carry}
\tilde{\rho} (\phi) = \frac{2 \lambda a^2(\phi )}{1-a^2(\phi)}
\qquad \tilde{a}(\phi) =\left( \frac{\rho}{\rho +2\lambda} 
\right)^{\ell /2}   ,
\end{equation}
where standard scalar field braneworlds are mapped onto one another 
if $\ell =1$ and a standard brane cosmology is dual to a phantom 
model for $\ell  =-1$. A further consequence of Eq. (\ref{carry}) 
is that the standard and phantom models generated from the 
same braneworld are related by a scale 
factor duality, $\tilde{a} (\phi ) \leftrightarrow 1/\hat{a} (\phi )$. 

In conclusion, a triality has been established between 
accelerating and decelerating cosmologies sourced by conventional and 
phantom scalar fields. The correspondence becomes apparent within the 
Hamilton--Jacobi framework of scalar field dynamics and arises in cosmological 
models based on Einstein gravity as well as the RSII braneworld.



\end{document}